\documentclass[twocolumn,notitlepage,superscriptaddress,showpacs,nofootinbib,tightenlines]{revtex4-1}
\usepackage{amsmath,verbatim,latexsym,amssymb,graphicx,indentfirst,mathrsfs,mathtools,amsthm,bbm,bm,hyperref,url,cancel,subcaption}
\usepackage[font=small,labelfont=bf,format=plain,justification=raggedright,singlelinecheck=false]{caption}

\usepackage[title,titletoc]{appendix}

\newtheoremstyle{indented}{5pt}{3pt}{\addtolength{\leftskip}{3.5em}}{}{\bfseries}{.}{.5em}{}
\theoremstyle{indented}
\newtheorem{theorem}{Theorem}

\makeatletter
\newcommand\footnoteref[1]{\protected@xdef\@thefnmark{\ref{#1}}\@footnotemark}
\makeatother

\renewcommand{\thesection}{\Roman{section}}
\renewcommand{\thesubsection}{\Roman{section}.\Alph{subsection}}
\renewcommand{\thesubsubsection}{\Roman{section}.\Alph{subsection}.\arabic{subsubsection}}
\makeatletter
\def\p@subsection{}
\makeatother
\makeatletter
\def\p@subsubsection{}
\makeatother

\usepackage{soul}

\usepackage{color}
\usepackage[dvipsnames]{xcolor}
\usepackage[normalem]{ulem}

\newcommand{\W}{ \mathcal{W} }  
\newcommand{\C}{ F }  
  
\newcommand{\Dim}{ d }  
 
\newcommand{\Protocol}{ \mathcal{P} }

\newcommand{\GW}{ G_\W }

\newcommand{\DegenW}{ \alpha }  
\newcommand{\DegenV}{ \lambda }  
\newcommand{\Charac}{ \mathcal{G} }  
\newcommand{\NondegW}{ \tilde{\W} } 
\newcommand{\NondegV}{ \tilde{V} } 
\newcommand{\U}{ \mathcal{U} } 
\newcommand{\Prob}{ \mathscr{P} } 
\newcommand{\TProb}{ \tilde{ \Prob } } 
\newcommand{\weak}{ {\rm weak} } 
\newcommand{\PWeak}{ \mathscr{P}_\weak } 
\newcommand{\WeakInt}{ \mathcal{I} }

\newcommand{\cc}{ {\rm c.c.} }

\newcommand{\Tr}{{\rm Tr}}   
\def\id{\mathbbm{1}}   
\newcommand{\kB}{k_\mathrm{B}}  

\newcommand{\LParen}{ \bm{(} }
\newcommand{\RParen}{ \bm{)} }
\newcommand*{\Set}[1]{\left\{  #1  \right\}}

\newcommand*{\ket}[1]{\lvert #1 \rangle}

\newcommand*{\ketbra}[2]{\lvert #1 \rangle\!\langle #2 \rvert}
\newcommand*{\expval}[1]{\left\langle  #1  \right\rangle}

\usepackage{graphicx}
\usepackage{epstopdf}
\newcommand{\caphead}[1]{{\bf #1}}

\begin{document}

\title{Jarzynski-like equality for the out-of-time-ordered correlator}
\author{Nicole~Yunger~Halpern\footnote{E-mail: nicoleyh@caltech.edu}}
\affiliation{Institute for Quantum Information and Matter, Caltech, Pasadena, CA 91125, USA}
\date{\today}


%
%

%
%
\begin{abstract}
The out-of-time-ordered correlator (OTOC) diagnoses quantum chaos
and the scrambling of quantum information
via the spread of entanglement.
The OTOC encodes forward and reverse evolutions
and has deep connections with the flow of time.
So do fluctuation relations such as Jarzynski's Equality,
derived in nonequilibrium statistical mechanics.
I unite these two powerful, seemingly disparate tools
by deriving a Jarzynski-like equality for the OTOC.
The equality's left-hand side equals the OTOC.
The right-hand side suggests a protocol 
for measuring the OTOC indirectly.
The protocol is platform-nonspecific and can be performed 
with weak measurement or with interference.
Time evolution need not be reversed in any interference trial.
The equality opens holography, condensed matter, and quantum information
to new insights from fluctuation relations and vice versa.
\end{abstract}
\maketitle{}

The out-of-time-ordered correlator (OTOC) $\C(t)$ 
diagnoses the scrambling of quantum information~\cite{Shenker_Stanford_14_BHs_and_butterfly,Shenker_Stanford_14_Multiple_shocks,Shenker_Stanford_15_Stringy,Roberts_15_Localized_shocks,Roberts_Stanford_15_Diagnosing,Maldacena_15_Bound}:
Entanglement can grow rapidly in a many-body quantum system,
dispersing information throughout many degrees of freedom.
$\C(t)$ quantifies the hopelessness of 
attempting to recover the information 
via local operations.

Originally applied to superconductors~\cite{LarkinO_69},
$\C(t)$ has undergone a revival recently.
$\C(t)$ characterizes quantum chaos, holography, black holes, and condensed matter.
The conjecture that black holes scramble quantum information
at the greatest possible rate
has been framed in terms of $\C(t)$~\cite{Maldacena_15_Bound,Sekino_Susskind_08_Fast_scramblers}.
The slowest scramblers include disordered systems~\cite{Huang_16_MBL_OTOC,Swingle_16_MBL_OTOC,Fan_16_MBL_OTOC,He_16_MBL_OTOC,Chen_16_MBL_OTOC}.
In the context of quantum channels,
$\C(t)$ is related to the tripartite information~\cite{HosurYoshida_16_Chaos}.
Experiments have been proposed~\cite{Swingle_16_Measuring,Yao_16_Interferometric,Zhu_16_Measurement}
and performed~\cite{Li_16_Measuring,Garttner_16_Measuring}
to measure $\C(t)$ with cold atoms and ions, 
with cavity quantum electrodynamics,
and with nuclear-magnetic-resonance quantum simulators.

$\C(t)$ quantifies sensitivity to initial conditions, a signature of chaos.
Consider a quantum system $S$ governed by a Hamiltonian $H$.
Suppose that $S$ is initialized to a pure state $\ket{ \psi }$
and perturbed with a local unitary operator $V$.
$S$ then evolves forward in time under the unitary
$U = e^{ - i H t}$ for a duration $t$,
is perturbed with a local unitary operator $\W$,
and evolves backward under $U^\dag$.
The state $\ket{ \psi' }  :=  U^\dag  \W  U  V  \ket{\psi}
= \W (t) V \ket{\psi}$ 
results.
Suppose, instead, that $S$ is perturbed with $V$
not at the sequence's beginning, but at the end:
$\ket{\psi}$ evolves forward under $U$,
is perturbed with $\W$,
evolves backward under $U^\dag$,
and is perturbed with $V$.
The state $\ket{ \psi'' }  :=  V U^\dag \W U \ket{ \psi } 
= V \W(t) \ket{\psi}$ results.
The overlap between the two possible final states equals the correlator:
$\C(t)  :=  \expval{ \W^\dag(t) \, V^\dag \, \W(t) \, V }  
=  \langle \psi'' | \psi' \rangle$.
The decay of $\C(t)$ reflects the growth
of $[\W(t), \, V]$~\cite{Maldacena_16_Comments,Polchinski_16_Spectrum}.

Forward and reverse time evolutions,
as well as information theory and diverse applications,  
characterize not only the OTOC, but also fluctuation relations.
Fluctuation relations have been derived 
in quantum and classical
nonequilibrium statistical mechanics~\cite{Jarzynski97,Crooks99,Tasaki00,Kurchan00}.
Consider a Hamiltonian $H(t)$
tuned from $H_i$ to $H_f$ at a finite speed.
For example, electrons may be driven within a circuit~\cite{Saira_12_Test}. 
Let $\Delta F  :=  F (H_f)  -  F( H_i )$
denote the difference between the equilibrium free energies
at the inverse temperature $\beta$:\footnote{
$F ( H_\ell )$ denotes the free energy in statistical mechanics,
while $F(t)$ denotes the OTOC in high energy and condensed matter.}
$F( H_\ell )  =  - \frac{1}{\beta} \ln Z_{\beta, \ell}$,
wherein the partition function is 
$Z_{\beta, \ell}  :=  \Tr ( e^{ - \beta H_\ell } )$
and $\ell = i, f$.
The free-energy difference
has applications in chemistry, biology, and pharmacology~\cite{Chipot_07_Free}.
One could measure $\Delta F$, in principle, by measuring
the work required to tune $H(t)$ from $H_i$ to $H_f$
while the system remains in equilibrium.
But such quasistatic tuning would require an infinitely long time.

$\Delta F$ has been inferred in a finite amount of time 
from Jarzynski's fluctuation relation,
$\expval{ e^{ - \beta W} }  =  e^{ - \beta \Delta F}$.
The left-hand side can be inferred from data about experiments
in which $H(t)$ is tuned from $H_i$ to $H_f$
arbitrarily quickly.
The work required to tune $H(t)$ during some particular trial
(e.g., to drive the electrons) is denoted by $W$.
$W$ varies from trial to trial 
because the tuning can eject the system arbitrarily far from equilibrium.
The expectation value $\langle \, . \, \rangle$ 
is with respect to the probability distribution $P(W)$
associated with any particular trial's
requiring an amount $W$ of work.
Nonequilibrium experiments have been combined
with fluctuation relations to estimate $\Delta F$~\cite{CollinRJSTB05,Douarche_05_Experimental,Blickle_06_Thermo,Harris_07_Experimental,MossaMFHR09,ManosasMFHR09,Saira_12_Test,Batalhao_14_Experimental,An_15_Experimental}:
\begin{align}
   \label{eq:DeltaF}
   \Delta F  =  - \frac{ 1 }{ \beta }  \:   
   \log  \expval{ e^{ - \beta W} }  \, .
\end{align}

Jarzynski's Equality, with the exponential's convexity, 
implies $\expval{ W }   \geq  \Delta F$.
The average work $\expval{W}$ required to tune $H(t)$
according to any fixed schedule
equals at least the work $\Delta F$ required to tune $H(t)$ quasistatically.
This inequality has been regarded as a manifestation of
the Second Law of Thermodynamics.
The Second Law governs information loss~\cite{Maruyama_09_Colloquium},
similarly to the OTOC's evolution.

I derive a Jarzynski-like equality, analogous to Eq.~\eqref{eq:DeltaF}, for $\C(t)$
(Theorem~\ref{theorem:OTOC_FT}).
The equality unites two powerful tools 
that have diverse applications
in quantum information, high-energy physics,
statistical mechanics, and condensed matter.
The union sheds new light on
both fluctuation relations and the OTOC,
similar to the light shed 
when fluctuation relations were introduced 
into ``one-shot'' statistical mechanics~\cite{Aberg_13_Truly,YungerHalpern_15_Introducing,Salek_15_Fluctuations,YungerHalpern_15_What,Dahlsten_15_Equality,Alhambra_16_Fluctuating_Work}.
The union also relates the OTOC,
known to signal quantum behavior in high energy and condensed matter,
to a quasiprobability,
known to signal quantum behavior in optics.
The Jarzynski-like equality suggests
a platform-nonspecific protocol for measuring $\C(t)$ indirectly.
The protocol can be implemented with weak measurements
or with interference.
The time evolution need not be reversed in any interference trial.
First, I present the set-up and definitions.
I then introduce and prove the Jarzynski-like equality for $\C(t)$.

\section{Set-up}

Let $S$ denote a quantum system
associated with a Hilbert space $\mathcal{H}$
of dimensionality $\Dim$.
The simple example of a spin chain~\cite{Yao_16_Interferometric,Zhu_16_Measurement,Li_16_Measuring,Garttner_16_Measuring} informs this paper:
Quantities will be summed over, as spin operators have discrete spectra.
Integrals replace the sums
if operators have continuous spectra.

Let $\W  =  \sum_{ w_\ell, \DegenW_{w_\ell} } w_\ell 
\ketbra{ w_\ell, \DegenW_{w_\ell} }{ w_\ell, \DegenW_{w_\ell} }$ and 
$V  =  \sum_{ v_\ell, \DegenV_{v_\ell} }  v_\ell  
\ketbra{ v_\ell, \DegenV_{v_\ell} }{ v_\ell, \DegenV_{v_\ell}}$ 
denote local unitary operators.
The eigenvalues are denoted by $w_\ell$ and $v_\ell$;
the degeneracy parameters, by $\DegenW_{w_\ell}$ and $\DegenV_{v_\ell}$.
$\W$ and $V$ may commute.
They need not be Hermitian.
Examples include single-qubit Pauli operators
localized at opposite ends of a spin chain.

We will consider measurements
of eigenvalue-and-degeneracy-parameter tuples
$(w_\ell, \DegenW_{w_\ell} )$ and $(v_\ell, \DegenV_{v_\ell} )$.
Such tuples can be measured as follows. 
A Hermitian operator
$\GW  =  \sum_{ w_\ell, \DegenW_{w_\ell} } 
   g ( w_\ell ) 
   \ketbra{ w_\ell, \DegenW_{w_\ell} }{ w_\ell, \DegenW_{w_\ell} }$ 
generates the unitary $\W$.
The generator's eigenvalues are labeled by the unitary's eigenvalues:
$w = e^{i g ( w_\ell ) }$.
Additionally, there exists a Hermitian operator
that shares its eigenbasis with $\W$
but whose spectrum is nondegenerate:
$\tilde{G}_{ \W }  =  \sum_{w_\ell, \DegenW_{w_\ell}}  
\tilde{g} ( \DegenW_{w_\ell} )
\ketbra{ w_\ell, \DegenW_{w_\ell} }{ w_\ell, \DegenW_{w_\ell} }$,
wherein $\tilde{g} ( \DegenW_{w_\ell} )$ denotes a real one-to-one function.
I refer to a collective measurement of $\GW$ and $\tilde{G}_{ \W }$
as a $\NondegW$ measurement.
Analogous statements concern $V$.
If $\Dim$ is large, measuring $\NondegW$ and $\NondegV$ 
may be challenging but is possible in principle.
Such measurements may be reasonable if $S$ is small.
Schemes for avoiding measurements of 
the $\DegenW_{w_\ell}$'s and $\DegenV_{v_\ell}$'s
are under investigation~\cite{BrianDisc}.

Let $H$ denote a time-independent Hamiltonian.
The unitary $U = e^{ - i H t}$ evolves $S$ forward in time for an interval $t$.
Heisenberg-picture operators are defined as
$\W(t) :=  U^\dag \W U$ and
$\W^\dag(t)  =  [ \W(t) ]^\dag  =  U^\dag \W^\dag U$.

The OTOC is conventionally evaluated on
a Gibbs state $e^{ - H / T } / Z$,
wherein $T$ denotes a temperature:
$\C(t)  =  \Tr \left( \frac{ e^{ - H / T } }{ Z } \W^\dag (t) V^\dag \W(t) V \right)$.
Theorem~\ref{theorem:OTOC_FT} generalizes beyond 
$e^{ - H / T } / Z$ to arbitrary density operators
$\rho = \sum_j  p_j  \ketbra{j}{j}  \in  \mathcal{D} ( \mathcal{H} )$.
[$\mathcal{D} ( \mathcal{H} )$ denotes the set of density operators
defined on $\mathcal{H}$.]

\section{Definitions}

Jarzynski's Equality concerns thermodynamic work, $W$.
$W$ is a random variable
calculated from measurement outcomes.
The out-of-time-ordering in $\C(t)$ requires two such random variables.
I label these variables $W$ and $W'$.

Two stepping stones connect $\W$ and $V$ to $W$ and $W'$.
First, I define a complex probability amplitude
$A_\rho ( w_2  , \DegenW_{w_2}  ;  v_1  ,  \DegenV_{v_1} ;
w_1  , \DegenW_{w_1}  ; j)$
associated with a quantum protocol.
I combine amplitudes $A_\rho$
into a $\tilde{A}_\rho$ inferable
from weak measurements and from interference.
$\tilde{A}_\rho$ resembles a quasiprobability,
a quantum generalization of a probability.
In terms of the $w_\ell$'s and $v_\ell$'s in $\tilde{A}_\rho$,
I define the measurable random variables $W$ and $W'$.

Jarzynski's Equality involves a probability distribution $P(W)$
over possible values of the work.
I define a complex analog $P(W, W')$.
These definitions are designed to parallel
expressions in~\cite{TLH_07_Work}.
Talkner, Lutz and H\"{a}nggi cast Jarzynski's Equality
in terms of a time-ordered correlation function.
Modifying their derivation will lead to the OTOC Jarzynski-like equality.

\subsection{Quantum probability amplitude $A_\rho$}

\begin{figure}[h]
\centering
\begin{subfigure}{0.4\textwidth}
\centering
\includegraphics[width=.9\textwidth]{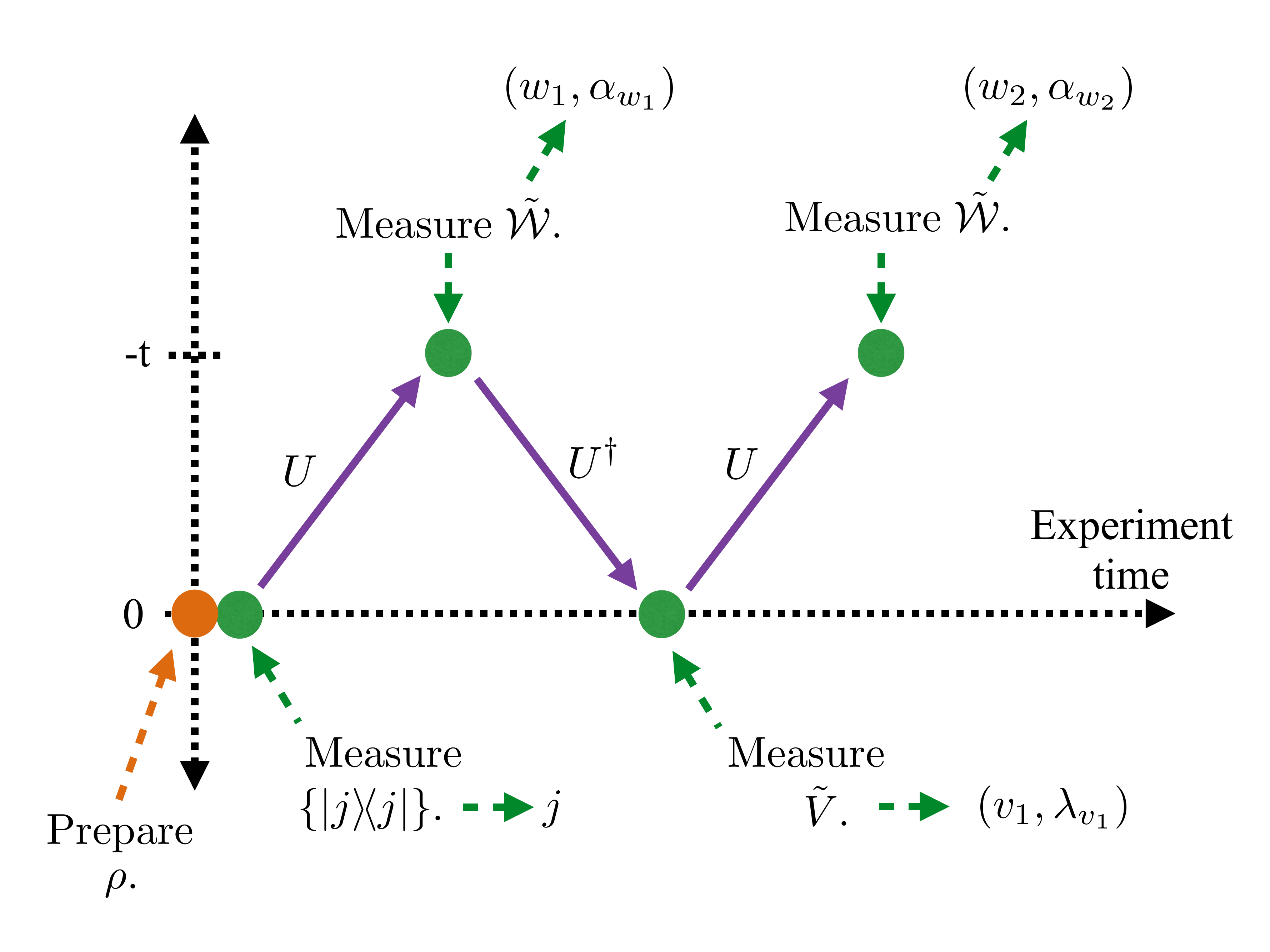}
\caption{}
\label{fig:Protocoll_Trial1}
\end{subfigure}
\begin{subfigure}{.4\textwidth}
\centering
\includegraphics[width=.9\textwidth]{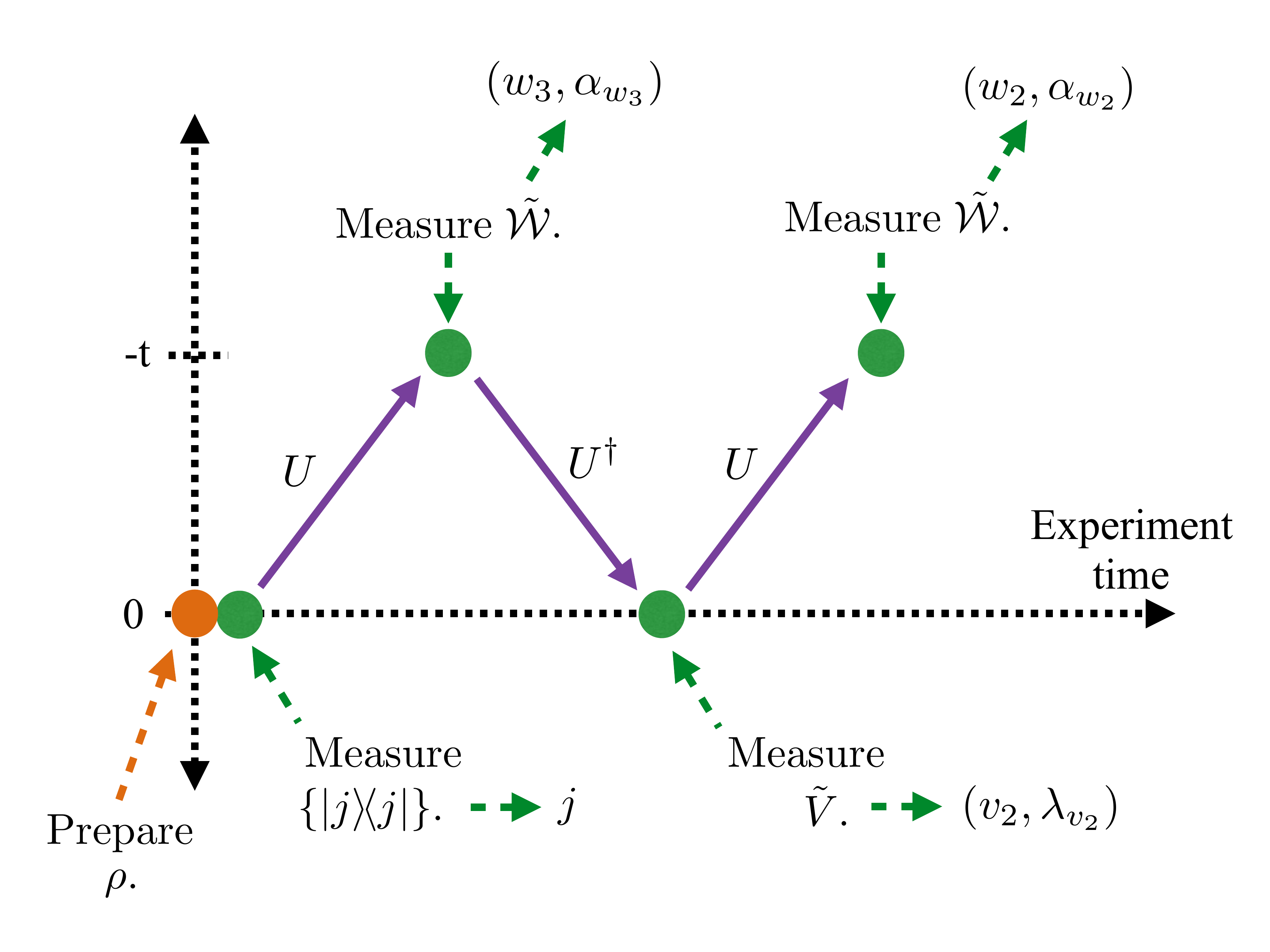}
\caption{}
\label{fig:Protocoll_Trial2}
\end{subfigure}
\caption{\caphead{Quantum processes described by 
the complex amplitudes in
the Jarzynski-like equality for the out-of-time-ordered correlator (OTOC):}
Theorem~\ref{theorem:OTOC_FT} shows that the OTOC
depends on a complex distribution $P(W, W')$.
This $P(W, W')$ parallels the probability distribution
over possible values of thermodynamic work 
in Jarzynski's Equality.
$P(W, W')$ results from summing products 
$A_\rho^*( . ) A_\rho( . )$.
Each $A_\rho( . )$ denotes a probability amplitude [Eq.~\eqref{eq:ADef}],
so each product resembles a probability.
But the amplitudes' arguments differ,
due to the OTOC's out-of-time ordering:
The amplitudes correspond to different quantum processes.
Figure~\ref{fig:Protocoll_Trial1} illustrates 
the process associated with the $A_\rho^*( . )$;
and Fig.~\ref{fig:Protocoll_Trial2}, the process associated with the $A_\rho( . )$.
Time runs from left to right.
Each process begins with the preparation of 
the state $\rho = \sum_j p_j \ketbra{j}{j}$ 
and a measurement of the state's eigenbasis.
Three evolutions ($U$, $U^\dag$, $U$) then alternate with 
three measurements of observables ($\NondegW$, $\NondegV$, $\NondegW$).
If the initial state commutes with the Hamiltonian $H$
(e.g., if $\rho = e^{ - H / T } / Z$),
the first $U$ can be omitted.
Figures~\ref{fig:Protocoll_Trial1} and~\ref{fig:Protocoll_Trial2}
are used to define $P(W, W')$,
rather than illustrating protocols for measuring $P(W, W')$.
$P(W, W')$ can be inferred from weak measurements
and from interferometry.}
\label{fig:Protocoll}
\end{figure}

The probability amplitude $A_\rho$
is defined in terms of the following protocol, $\Protocol$:
\begin{enumerate}
   \item Prepare $\rho$.
   \item Measure the eigenbasis of $\rho$, $\{ \ketbra{j}{j} \}$. 
   \item \label{item:FirstU}
            Evolve $S$ forward in time under $U$.
   \item Measure $\NondegW$. 
   \item Evolve $S$ backward in time under $U^\dag$.
   \item Measure $\NondegV$. 
   \item Evolve $S$ forward under $U$.
   \item Measure $\NondegW$. 
\end{enumerate}
An illustration appears in Fig.~\ref{fig:Protocoll_Trial1}.
Consider implementing $\Protocol$ in one trial.
The complex probability amplitude
associated with the measurements' yielding $j$,
then $( w_1, \DegenW_{w_1} )$, then $(v_1, \DegenV_{v_1} )$, 
then $( w_2, \DegenW_{w_2} )$ is
\begin{align}
   \label{eq:ADef}
   & A_\rho( w_2  , \DegenW_{w_2}  ;  v_1  ,  \DegenV_{v_1} ;
w_1  , \DegenW_{w_1}  ; j)
   :=  \langle w_2, \DegenW_{w_2} | U | v_1, \DegenV_{v_1} \rangle
   \nonumber \\ & \qquad \times
   \langle v_1, \DegenV_{v_1}  |  U^\dag  |  w_1, \DegenW_{w_1}  \rangle
   \langle w_1,  \DegenW_{w_1}  |  U | j \rangle
   \sqrt{ p_j } \, .
\end{align}
The square modulus $ | A_\rho ( . ) |^2$ equals
the joint probability that these measurements yield these outcomes.

Suppose that $[\rho, \, H] = 0$.
For example, suppose that 
$S$ occupies the thermal state $\rho = e^{ - H / T } / Z$.
(I set Boltzmann's constant to one: $\kB = 1$.)
Protocol $\Protocol$ and Eq.~\eqref{eq:ADef} simplify:
The first $U$ can be eliminated,
because $[\rho, \, U] = 0$.
Why $[\rho, \, U] = 0$ obviates the unitary
will become apparent when 
we combine $A_\rho$'s into $\tilde{A}_\rho$.

The protocol $\Protocol$ defines $A_\rho$;
$\Protocol$ is not a prescription measuring $A_\rho$.
Consider implementing $\Protocol$ many times
and gathering statistics about the measurements' outcomes.
From the statistics, one can infer
the probability $| A_\rho |^2$,
not the probability amplitude $A_\rho$.
$\Protocol$ merely is the process 
whose probability amplitude equals $A_\rho$.
One must calculate combinations of $A_\rho$'s
to calculate the correlator.
These combinations, labeled $\tilde{A}_\rho$,
can be inferred from weak measurements and interference.

%
%
\subsection{Combined quantum amplitude $\tilde{A}_\rho$}
\label{section:ADirac}

Combining quantum amplitudes $A_\rho$ 
yields a quantity $\tilde{A}_\rho$
that is nearly a probability
but that differs due to the OTOC's out-of-time ordering.
I first define $\tilde{A}_\rho$, 
which resembles the Kirkwood-Dirac quasiprobability~\cite{Kirkwood_33_Quantum,Dirac_45_On,Dressel_15_Weak,BrianDisc}.
We gain insight into $\tilde{A}_\rho$ 
by supposing that $[\rho, \, \W] = 0$,
e.g., that $\rho$ is the infinite-temperature Gibbs state $\id / \Dim$.
$\tilde{A}_\rho$ can reduce to a probability in this case,
and protocols for measuring $\tilde{A}_\rho$ simplify.
I introduce weak-measurement and interference schemes
for inferring $\tilde{A}_\rho$ experimentally.

\subsubsection{Definition of the combined quantum amplitude $\tilde{A}_\rho$}
\label{section:TildeA}

Consider measuring the probability amplitudes $A_\rho$
associated with all the possible measurement outcomes.
Consider fixing an outcome septuple 
$( w_2  , \DegenW_{w_2}  ;  v_1  ,  \DegenV_{v_1} ;
w_1  , \DegenW_{w_1}  ; j)$.
The amplitude $A_\rho( w_2  , \DegenW_{w_2}  ;  v_1  ,  \DegenV_{v_1} ;
w_1  , \DegenW_{w_1}  ; j)$
describes one realization, illustrated in Fig.~\ref{fig:Protocoll_Trial1},
of the protocol $\Protocol$.
Call this realization $a$.

Consider the $\Protocol$ realization, labeled $b$,
illustrated in Fig.~\ref{fig:Protocoll_Trial2}.
The initial and final measurements
yield the same outcomes as in $a$
[outcomes $j$ and $(w_2, \DegenW_{w_2} )$].
Let $(w_3, \DegenW_{w_3} )$ and $( v_2, \DegenV_{v_2} )$
denote the outcomes of
the second and third measurements in $b$.
Realization $b$ corresponds to the probability amplitude
$A_\rho( w_2  , \DegenW_{w_2}  ;  v_2  ,  \DegenV_{v_2} ;
w_3  , \DegenW_{w_3}  ; j)$.

Let us complex-conjugate the $b$ amplitude
and multiply by the $a$ amplitude.
We marginalize over $j$ and over $(w_1,  \DegenW_{w_1} )$,
forgetting about the corresponding measurement outcomes:
\begin{align}
   \label{eq:TildeADef}
   \tilde{A}_\rho ( w, v, \DegenW_w, \DegenV_v )   & :=  
   \sum_{ j,  (w_1,  \DegenW_{w_1} ) }
   A^*_\rho( w_2  , \DegenW_{w_2}  ;  v_2  ,  \DegenV_{v_2} ;
w_3  , \DegenW_{w_3}  ; j)
   \nonumber \\ &  \; \times
   A_\rho( w_2  , \DegenW_{w_2}  ;  v_1  ,  \DegenV_{v_1} ;
w_1  , \DegenW_{w_1}  ; j ) \, .
\end{align}
The shorthand $w$ encapsulates the list $(w_1, w_2)$.
The shorthands $v$, $\DegenW_w$ and $\DegenV_v$
are defined analogously. 

Let us substitute in from Eq.~\eqref{eq:ADef}
and invoke $\langle A | B \rangle^*  =  \langle B | A \rangle$.
The sum over $(w_1,  \DegenW_{w_1} )$ evaluates
to a resolution of unity.
The sum over $j$ evaluates to $\rho$:
\begin{align}
   \label{eq:TildeAExp}
   & \tilde{A}_\rho ( w, v, \DegenW_w, \DegenV_v )
   =    \langle  w_3,  \DegenW_{w_3}  |  U  |  v_2,  \DegenV_{v_2}  \rangle
   \langle  v_2,  \DegenV_{v_2}  |  U^\dag  |  w_2,  \DegenW_{w_2}  \rangle
   \nonumber \\ & \qquad \times
   \langle  w_2,  \DegenW_{w_2}  |  U  |  v_1,  \DegenV_{v_1}  \rangle
   \langle  v_1,  \DegenV_{v_1}  |  \rho  U^\dag  |  w_3,  \DegenW_{w_3}  \rangle   \, .
\end{align}

This $\tilde{A}_\rho$ resembles the Kirkwood-Dirac quasiprobability~\cite{Dressel_15_Weak,BrianDisc}.
Quasiprobabilities surface in quantum optics and quantum foundations~\cite{Carmichael_02_Statistical,Ferrie_11_Quasi}.
Quasiprobabilities generalize probabilities to quantum settings.
Whereas probabilities remain between 0 and 1,
quasiprobabilities can assume negative and nonreal values.
Nonclassical values signal quantum phenomena such as entanglement.
The best-known quasiprobabilities include
the Wigner function, the Glauber-Sudarshan $P$ representation,
and the Husimi $Q$ representation.
Kirkwood and Dirac defined another quasiprobability
in 1933 and in 1945~\cite{Kirkwood_33_Quantum,Dirac_45_On}.
Interest in the Kirkwood-Dirac quasiprobability has revived recently.
The distribution can assume nonreal values,
obeys Bayesian updating, and has been measured experimentally~\cite{Lundeen_11_Direct,Lundeen_12_Procedure,Bamber_14_Observing,Mirhosseini_14_Compressive}.

The Kirkwood-Dirac distribution for a state 
$\sigma \in \mathcal{D}( \mathcal{H} )$ has the form
$\langle f | a \rangle   \langle a |  \sigma  | f \rangle$,
wherein $\Set{ \ketbra{f}{f} }$ and $\Set{ \ketbra{a}{a} }$
denote bases for $\mathcal{H}$~\cite{Dressel_15_Weak}.
Equation~\eqref{eq:TildeAExp} has the same form
except contains more outer products.
Marginalizing $\tilde{A}_\rho$ over every variable
except one $w_\ell$ [or one $v_\ell$, 
one $( w_\ell,  \, \DegenW_{w_\ell} )$,
or one $( v_\ell,  \,  \DegenV_{v_\ell} )$]
yields a probability,
as does marginalizing the Kirkwood-Dirac distribution
over every variable except one.
The precise nature of the relationship between 
$\tilde{A}_\rho$ and the Kirkwood-Dirac quasiprobability
is under investigation~\cite{BrianDisc}.
For now, I harness the similarity 
to formulate a weak-measurement scheme for $\tilde{A}_\rho$
in Sec.~\ref{section:WeakMain}.

$\tilde{A}_\rho$ is nearly a probability:
$\tilde{A}_\rho$ results from multiplying
a complex-conjugated probability amplitude $A^*_\rho$
by a probability amplitude $A_\rho$.
So does the quantum mechanical probability density
$p(x) = \psi^*(x) \psi(x)$.
Hence the quasiprobability resembles a probability.
Yet the argument of the $\psi^*$ equals
the argument of the $\psi$.
The argument of the $A^*_\rho$ does not equal
the argument of the $A_\rho$.
This discrepancy stems from the OTOC's out-of-time ordering.
$\tilde{A}_\rho$ can be regarded
as like a probability,
differing due to the out-of-time ordering.
$\tilde{A}_\rho$ reduces to a probability 
under conditions discussed in Sec.~\ref{section:SimpleTildeA}.
The reduction reinforces the parallel between
Theorem~\ref{theorem:OTOC_FT} and
the fluctuation-relation work~\cite{TLH_07_Work},
which involves a probability distribution that resembles $\tilde{A}_\rho$.

\subsubsection{Simple case, 
reduction of $\tilde{A}_\rho$ to a probability} 
\label{section:SimpleTildeA}

Suppose that $\rho$ shares the $\NondegW(t)$ eigenbasis:
$\rho  =  \rho_{ \W(t) }
:=  \sum_{w_\ell, \DegenW_{w_\ell} }  p_{ w_\ell, \DegenW_{w_\ell} }
U^\dag   \ketbra{ w_\ell, \DegenW_{w_\ell} }{ w_\ell, \DegenW_{w_\ell} }   U$.
For example, $\rho$ may be 
the infinite-temperature Gibbs state $\id / \Dim$.
Equation~\eqref{eq:TildeAExp} becomes
\begin{align}
   \label{eq:TildeASimple}
   & \tilde{A}_{ \rho_{ \W(t) } } ( w, v, \DegenW_w,  \DegenV_v )
   =  \langle w_3, \DegenW_{w_3}  |  U  |  v_2, \DegenV_{v_2}  \rangle
   \nonumber \\  & \qquad \times
   \langle v_2, \DegenV_{v_2}  |  U^\dag  |  w_2,  \DegenW_{w_2}  \rangle
   \langle  w_2,  \DegenW_{w_2}  |  U  |  v_1,  \DegenV_{v_1}  \rangle
   \nonumber \\  & \qquad \times
   \langle  v_1,  \DegenV_{v_1}  |  U^\dag  |  w_3, \DegenW_{w_3}  \rangle
   \, p_{w_3, \DegenW_{w_3} } \, .
\end{align}
The weak-measurement protocol simplifies,
as discussed in Sec.~\ref{section:WeakMain}.

Equation~\eqref{eq:TildeASimple}
reduces to a probability if 
$(w_3, \DegenW_{w_3})  =  ( w_2,  \DegenW_{w_2})$ or if
$( v_2,  \DegenV_{v_2} )  =  ( v_1,  \DegenV_{v_1} )$.
For example, suppose that
$(w_3, \DegenW_{w_3})  =  ( w_2,  \DegenW_{w_2})$:
\begin{align}
   \label{eq:TildeASimple2}
   & \tilde{A}_{ \rho_{ \W(t) } } \LParen  
   ( w_2, w_2 ),  v ,  ( \DegenW_{w_2} ,  \DegenW_{w_2} ),  \DegenV_v  \RParen
   = | \langle v_2,  \DegenV_{v_2}  | U^\dag | w_2,  \DegenW_{w_2}  \rangle |^2
   \nonumber \\ & \qquad \qquad \qquad  \times
   | \langle  v_1,  \DegenV_{v_1}  |  U^\dag  |  w_2,  \DegenW_{w_2}  \rangle  |^2 \,
   p_{ w_2, \DegenW_{w_2} } \\
   \label{eq:ReduceToProb}
    & \qquad = 
    p ( v_2, \DegenV_{v_2} | w_2, \DegenW_{w_2} ) \,
    p ( v_1, \DegenV_{v_1} | w_2, \DegenW_{w_2} ) \,
    p_{ w_2, \DegenW_{w_2} } \, .
\end{align}
The $p_{ w_2, \DegenW_{w_2} }$ denotes the probability that
preparing $\rho$ and measuring $\NondegW$
will yield $(w_2, \DegenW_{w_2})$.
Each $p( v_\ell, \DegenV_{v_\ell} | w_2, \DegenW_{w_2} )$ denotes
the conditional probability that
preparing $\ket{w_2, \DegenW_{w_2} }$, backward-evolving under $U^\dag$, 
and measuring $\NondegV$ 
will yield $(v_\ell, \DegenV_{v_\ell})$.
Hence the combination $\tilde{A}_\rho$ of probability amplitudes
is nearly a probability:
$\tilde{A}_\rho$ reduces to a probability under simplifying conditions.

Equation~\eqref{eq:ReduceToProb} strengthens the analogy 
between Theorem~\ref{theorem:OTOC_FT} and
the fluctuation relation in~\cite{TLH_07_Work}.
Equation~(10) in~\cite{TLH_07_Work} contains 
a conditional probability $p( m, t_f | n )$
multiplied by a probability $p_n$. 
These probabilities parallel the $p ( v_1, \DegenV_{v_1} | w_1, \DegenW_{w_1} )$
and $p_{ w_1, \DegenW_{w_1} }$ in Eq.~\eqref{eq:ReduceToProb}.
Equation~\eqref{eq:ReduceToProb} contains another conditional probability,
$p ( v_2, \DegenV_{v_2} | w_1, \DegenW_{w_1} )$,
due to the OTOC's out-of-time ordering.

\subsubsection{Weak-measurement scheme for 
the combined quantum amplitude $\tilde{A}_\rho$}
\label{section:WeakMain}

$\tilde{A}_\rho$ is related to the Kirkwood-Dirac quasiprobability,
which has been inferred from weak measurements~\cite{Dressel_14_Understanding,Kofman_12_Nonperturbative,Lundeen_11_Direct,Lundeen_12_Procedure,Bamber_14_Observing,Mirhosseini_14_Compressive}.
I sketch a weak-measurement scheme for inferring $\tilde{A}_\rho$.
Details appear in Appendix~\ref{section:WeakMeas}.

Let $\Protocol_\weak$ denote the following protocol:
\begin{enumerate}
   \item Prepare $\rho$.
   \item \label{item:FirstVWeak}
            Couple the system's $\NondegV$ weakly to an ancilla $\mathcal{A}_{ a }$. 
            Measure $\mathcal{A}_{ a }$ strongly.
   \item \label{item:FirstUWeak}
            Evolve $S$ forward under $U$.
   \item Couple the system's $\NondegW$ weakly to an ancilla $\mathcal{A}_{ b }$. 
            Measure $\mathcal{A}_{ b }$ strongly.
   \item Evolve $S$ backward under $U^\dag$.
   \item Couple the system's $\NondegV$ weakly to an ancilla $\mathcal{A}_c$.
            Measure $\mathcal{A}_c$ strongly.
   \item Evolve $S$ forward under $U$.
   \item Measure $\NondegW$ strongly (e.g., projectively). 
\end{enumerate}
Consider performing $\Protocol_\weak$ many times.
From the measurement statistics, one can infer the form of 
$\tilde{A}_\rho ( w, v, \DegenW_w, \DegenV_v )$.

$\Protocol_\weak$ offers an experimental challenge:
Concatenating weak measurements raises
the number of trials required to infer a quasiprobability.
The challenge might be realizable 
with modifications to existing set-ups (e.g.,~\cite{White_16_Preserving,Dressel_14_Implementing}).
Additionally, $\Protocol_\weak$ simplifies
in the case discussed in Sec.~\ref{section:SimpleTildeA}---if
$\rho$ shares the $\NondegW(t)$ eigenbasis,
e.g., if $\rho = \id / \Dim$.
The number of weak measurements reduces from three to two.
Appendix~\ref{section:WeakMeas} contains details.

\subsubsection{Interference-based measurement of $\tilde{A}_\rho$}

$\tilde{A}_\rho$ can be inferred
not only from weak measurement, but also from interference.
In certain cases---if $\rho$ shares neither 
the $\W$, the $\W(t)$, nor the $V$  eigenbasis---also
quantum state tomography is needed.
From interference, one infers the inner products 
$\langle a | \U | b \rangle$ in $\tilde{A}_\rho$.
Eigenstates of $\NondegW$ and $\NondegV$
are labeled by $a$ and $b$;
and $\U = U, U^\dag$.
The matrix element 
$\langle v_1, \DegenV_{v_1}  |  \rho  U^\dag  
|  w_3,  \DegenW_{w_3}  \rangle$
is inferred from quantum state tomography in certain cases.

The interference scheme proceeds as follows.
An ancilla $\mathcal{A}$ is prepared in
a superposition $\frac{1}{ \sqrt{2} } \; ( \ket{0} + \ket{1} )$.
The system $S$ is prepared in a fiducial state $\ket{f}$.
The ancilla controls a conditional unitary on $S$:
If $\mathcal{A}$ is in state $\ket{0}$, 
$S$ is rotated to $\U \ket{b}$.
If $\mathcal{A}$ is in $\ket{1}$, $S$ is rotated to $\ket{a}$.
The ancilla's state is rotated about
the $x$-axis [if the imaginary part $\Im ( \langle a | \U | b \rangle )$
is being inferred]
or about the $y$-axis [if the real part
$\Re ( \langle a | \U | b \rangle )$ is being inferred].
The ancilla's $\sigma_z$ and the system's $\Set{ \ket{a} }$ are measured.
The outcome probabilities imply the value of $\langle a | \U | b \rangle$.
Details appear in Appendix~\ref{section:Interfere}.

The time parameter $t$ need not be negated
in any implementation of the protocol.
The absence of time reversal has been regarded as beneficial
in OTOC-measurement schemes~\cite{Yao_16_Interferometric,Zhu_16_Measurement},
as time reversal can be difficult to implement.

Interference and weak measurement have been performed with cold atoms~\cite{Smith_04_Continuous},
which have been proposed as platforms for realizing scrambling and quantum chaos~\cite{Swingle_16_Measuring,Yao_16_Interferometric,Danshita_16_Creating}.
Yet cold atoms are not necessary for measuring $\tilde{A}_\rho$.
The measurement schemes in this paper are platform-nonspecific.

\subsection{Measurable random variables $W$ and $W'$}

The combined quantum amplitude $\tilde{A}_\rho$
is defined in terms of 
two realizations of the protocol $\Protocol$.
The realizations yield measurement outcomes
$w_2$, $w_3$, $v_1$, and $v_2$.
Consider complex-conjugating two outcomes:
$w_3  \mapsto  w^*_3$, and $v_2  \mapsto  v^*_2$.
The four values are combined into
\begin{align}
   \label{eq:WDefs}
   W := w_3^*  v_2^*
   \quad \text{and} \quad
   W'  :=  w_2 v_1 \, .
\end{align}

Suppose, for example, that $\W$ and $V$ denote
single-qubit Paulis. $( W, W')$ can equal
$( 1, 1 ) , ( 1, - 1 ) , ( -1, 1 )$, or $( -1, -1 )$.
$W$ and $W'$ function analogously to 
the thermodynamic work in Jarzynski's Equality:
$W$, $W'$, and work are random variables
calculable from measurement outcomes.

%
%
\subsection{Complex distribution function $P(W, W')$}
\label{section:PW}

Jarzynski's Equality depends on a probability distribution $P(W)$.
I define an analog $P(W, W')$
in terms of the combined quantum amplitude $\tilde{A}_\rho$.

Consider fixing $W$ and $W'$.
For example, let $(W, W' )  =  ( 1 , -1 )$.
Consider the set of all possible outcome octuples
$(w_2,  \DegenW_{w_2} ; w_3, \DegenW_{w_3} ;
    v_1 ,  \DegenV_{v_1} ;  v_2, \DegenV_{v_2}  )$
that satisfy the constraints $W = w_3^*  v_2^*$ and $W' = w_2 v_1$.
Each octuple corresponds to 
a set of combined quantum amplitudes
$\tilde{A}_\rho ( w, v, \DegenW_w, \DegenV_v )$.
These $\tilde{A}_\rho$'s are summed,
subject to the constraints:
\begin{align}
   \label{eq:PDef}
   P (W, W')  & :=  \sum_{ w, v, \DegenW_w, \DegenV_v }
   \tilde{A}_\rho  ( w, v, \DegenW_w, \DegenV_v )  
   \nonumber \\ & \qquad \times
   \delta_{ W ( w_3^*  v_2^* ) } \,
   \delta_{ W' ( w_2 v_1 ) } \, .
\end{align}
The Kronecker delta is denoted by $\delta_{ab}$.

The form of Eq.~\eqref{eq:PDef} is analogous to 
the form of the $P(W)$ in~\cite{TLH_07_Work} [Eq.~(10)],
as $\tilde{A}_\rho$ is nearly a probability.
Equation~\eqref{eq:PDef}, however, encodes interference
of quantum probability amplitudes.

$P(W, W')$ resembles a joint probability distribution.
Summing any function $f(W, W')$ with weights $P(W, W')$ 
yields the average-like quantity
\begin{align}
   \label{eq:ExpValDef}
   \expval{ f(W, W') }  :=
   \sum_{W, W'}  f(W, W')  \,  P(W, W') \, .
\end{align}

\section{Result} 

The above definitions feature in the Jarzynski-like equality for the OTOC.

\begin{theorem}
\label{theorem:OTOC_FT}
The out-of-time-ordered correlator
obeys the Jarzynski-like equality
\begin{align}
   \label{eq:Result}
   \C(t) =  \frac{ \partial^2 }{ \partial \beta  \,  \partial \beta' }  \:
      \expval{ e^{ - ( \beta W + \beta' W' ) }  }  
      \Big\lvert_{ \beta,  \beta' = 0 }   \, ,
\end{align}
wherein $\beta, \beta'  \in  \mathbb{R}$.
\end{theorem}
\noindent 

\begin{proof}

The derivation of Eq.~\eqref{eq:Result} 
is inspired by~\cite{TLH_07_Work}.
Talkner \emph{et al.} cast Jarzynski's Equality in terms of 
a time-ordered correlator of two exponentiated Hamiltonians.
Those authors invoke the characteristic function
\begin{align}
   \label{eq:CharDef1}
   \Charac(s)  :=  \int dW \; e^{i s W} \, P(W) \, ,
\end{align}
the Fourier transform of the probability distribution $P(W)$.
The integration variable $s$ is regarded as
an imaginary inverse temperature: $i s = - \beta$.
We analogously invoke the (discrete) Fourier transform of $P(W, W')$:
\begin{align}
   \label{eq:CharDef2}
   \Charac(s, s')  :=  \sum_W e^{i s W}   \sum_{W'} e^{i s' W'}
   P(W, W') \, ,
\end{align}
wherein $is = - \beta$ and $is'  =  - \beta'$.

$P(W, W')$ is substituted in from 
Eqs.~\eqref{eq:PDef} and~\eqref{eq:TildeAExp}.
The delta functions are summed over:
\begin{align}
   & \Charac (s, s')  =  
   \sum_{ w, v, \DegenW_w, \DegenV_v  }   
   e^{i s w_3^* v_2^*} \: e^{i s' w_2 v_1 } \:
   \langle  w_3,  \DegenW_{w_3}  |  U  |  v_2, \DegenV_{v_2}  \rangle
   \nonumber \\ & \qquad \times
   \langle  v_2,  \DegenV_{v_2}  |  U^\dag  |  w_2 ,  \DegenW_{w_2}  \rangle
   \langle     w_2 ,  \DegenW_{w_2}   |  U  |  v_1,  \DegenV_{v_1}  \rangle
   \nonumber \\ & \qquad \times
   \langle  v_1,  \DegenV_{v_1}  |  U^\dag  \rho(t)  |  w_3,  \DegenW_{w_3}  \rangle  \, .
\end{align}
The $\rho U^\dag$ in Eq.~\eqref{eq:TildeAExp} 
has been replaced with $U^\dag \rho(t)$,
wherein $\rho(t)  :=  U \rho U^\dag$.

The sum over $( w_3 ,  \DegenW_{w_3 }  )$ is recast as a trace.
Under the trace's protection, $\rho(t)$ is shifted
to the argument's left-hand side.
The other sums and the exponentials
are distributed across the product:
\begin{align}
   \Charac  & (s, s')  =
   \Tr \Bigg(  \rho(t) 
   \Bigg[ \sum_{ w_3, \DegenW_{w_3} }   
             \ketbra{ w_3, \DegenW_{w_3} }{ w_3, \DegenW_{w_3} }
             \nonumber \\ & \qquad \times
             U  \sum_{v_2, \DegenV_{v_2} }  
                      e^{is { {w_3}^* } v_2^*}    
                      \ketbra{ v_2, \DegenV_{v_2} }{ v_2, \DegenV_{v_2} }  U^\dag \Bigg]
   \nonumber \\ &  \times
   \Bigg[ \sum_{ w_2, \DegenW_{w_2} }  
             \ketbra{ w_2, \DegenW_{w_2} }{ w_2, \DegenW_{w_2} } 
             \nonumber \\ & \qquad \times
             U  \sum_{ v_1, \DegenV_{v_1} } 
                      e^{ i s' w_2 v_1 }  
                      \ketbra{  v_1, \DegenV_{v_1}  }{  v_1, \DegenV_{v_1}  }  \,  U^\dag
   \Bigg] \Bigg)  \, .
\end{align}

The $v_\ell$ and $\DegenV_{v_\ell}$ sums are eigendecompositions
of exponentials of unitaries:
\begin{align}
   \Charac  & (s, s')  =
   \Tr \Bigg(   \rho(t) 
   \Bigg[  \sum_{ w_3, \DegenW_{w_3} }   
             \ketbra{ w_3, \DegenW_{w_3} }{ w_3, \DegenW_{w_3} }     
             U  \,  e^{is w_3^* V^\dag}  \, U^\dag   \Bigg]
   \nonumber \\ & \qquad   \times
   \Bigg[  \sum_{ w_2,  \DegenW_{w_2} }  
              \ketbra{ w_2,  \DegenW_{w_2} }{ w_2,  \DegenW_{w_2} }   
              U  \,e^{ i s' w_2 V}  \, U^\dag  \Bigg]
   \Bigg)  \, .
\end{align}
The unitaries time-evolve the $V$'s:
\begin{align}
   \Charac  & (s, s')  =
   \Tr \Bigg(   \rho(t)
   \Bigg[  \sum_{  w_3, \DegenW_{w_3}  }   
            \ketbra{ w_3, \DegenW_{w_3} }{ w_3, \DegenW_{w_3} }   
            e^{is w_2^* V^\dag( -t )}  \Bigg]
   \nonumber \\ &   \qquad \times
   \Bigg[  \sum_{ w_2,  \DegenW_{w_2} }  
              \ketbra{ w_2,  \DegenW_{w_2} }{ w_2,  \DegenW_{w_2} }   
              e^{ i s' w_2 V( -t )}  \Bigg]  
   \Bigg)  \, .
\end{align}

We differentiate with respect to $is' = - \beta'$
and with respect to $is = - \beta$.
Then, we take the limit as $\beta, \beta' \to 0$:
\begin{align}
   \label{eq:GEq}
   & \frac{ \partial^2 }{ \partial \beta \, \partial \beta' }
   \Charac  \left( i \beta,  i \beta'  \right)  
   \Big\lvert_{\beta, \beta' = 0}
   \\  & 
   =   \Tr \Bigg(   \rho(t)
   \Bigg[  \sum_{  w_3, \DegenW_{w_3} }    w_3^* 
              \ketbra{ w_3, \DegenW_{w_3} }{ w_3, \DegenW_{w_3} }  
              V^\dag( -t )  \Bigg]
   \\ \nonumber & \qquad \qquad \times
   \Bigg[  \sum_{ w_2,  \DegenW_{w_2} }  w_2  
              \ketbra{ w_2,  \DegenW_{w_2} }{ w_2,  \DegenW_{w_2} }
              V( -t )  \Bigg]  \Bigg)  \\
   &  =   \label{eq:GEq2}
    \Tr \LParen  \rho(t)  \,
    \W^\dag  \,  V^\dag(-t)  \,  \W  \,  V( -t )  \RParen  \, .
\end{align}

Recall that $\rho(t)  :=  U \rho U^\dag$.
Time dependence is transferred 
from $\rho(t)$, $V(-t) = U V^\dag U^\dag$, 
and $V^\dag(t) = U V U^\dag$ to $\W^\dag$ and $\W$,
under the trace's cyclicality:
\begin{align}
   & \frac{ \partial^2 }{ \partial \beta \, \partial \beta' }
   \Charac  ( i \beta,  i \beta')  \Big\lvert_{\beta, \beta' = 0}
   \label{eq:GEq3}
   =   \Tr \left(  \rho  \,
   \W^\dag(t)  \,  V^\dag  \,  \W(t)  \,  V  \right)  \\
   & \qquad \qquad \qquad \qquad
   =  \expval{ \W^\dag(t)  \,  V^\dag  \,  \W(t)  \,  V } 
   =  \C(t) \, .
\end{align}
By Eqs.~\eqref{eq:ExpValDef} and~\eqref{eq:CharDef2},
the left-hand side equals 
\begin{align}
   \frac{ \partial^2 }{ \partial \beta \, \partial \beta' }
   \expval{ e^{ - ( \beta W + \beta' W') } }  
   \Big\lvert_{\beta, \beta' = 0} \, .
\end{align}
\end{proof} 

Theorem~\ref{theorem:OTOC_FT} resembles 
Jarzynski's fluctuation relation in several ways.
Jarzynski's Equality encodes a scheme for measuring 
the difficult-to-calculate $\Delta F$ from realizable nonequilibrium trials.
Theorem~\ref{theorem:OTOC_FT} encodes a scheme for measuring
the difficult-to-calculate $\C(t)$ from realizable nonequilibrium trials.
$\Delta F$ depends on just 
a temperature and two Hamiltonians.
Similarly, the conventional $\C(t)$ 
(defined with respect to $\rho = e^{ - H / T } / Z$) 
depends on just a temperature, a Hamiltonian, 
and two unitaries.
Jarzynski relates $\Delta F$ to 
the characteristic function of a probability distribution.
Theorem~\ref{theorem:OTOC_FT} relates $\C(t)$ to
(a moment of) the characteristic function
of a (complex) distribution.

The complex distribution, $P(W, W')$,
is a combination of probability amplitudes $\tilde{A}_\rho$
related to quasiprobabilities.
The distribution in Jarzynski's Equality
is a combination of probabilities.
The quasiprobability-vs.-probability contrast  
fittingly arises from the OTOC's out-of-time ordering.
$\C(t)$ signals quantum behavior (noncommutation), 
as quasiprobabilities signal quantum behaviors (e.g., entanglement).
Time-ordered correlators similar to $\C(t)$ 
track only classical behaviors
and are moments of (summed) classical probabilities~\cite{BrianDisc}.
OTOCs that encode more time reversals than $\C(t)$
are moments of combined quasiprobability-like distributions
lengthier than $\tilde{A}_\rho$~\cite{BrianDisc}.

\section{Conclusions}

The Jarzynski-like equality for the out-of-time correlator
combines an important tool from nonequilibrium statistical mechanics
with an important tool from 
quantum information, high-energy theory, and condensed matter.
The union opens all these fields to new modes of analysis.

For example, Theorem~\ref{theorem:OTOC_FT} relates the OTOC
to a combined quantum amplitude $\tilde{A}_\rho$.
This $\tilde{A}_\rho$ is closely related to a quasiprobability.
The OTOC and quasiprobabilities have signaled nonclassical behaviors
in distinct settings---in high-energy theory and condensed matter 
and in quantum optics, respectively.
The relationship between OTOCs and quasiprobabilities merits study:
What is the relationship's precise nature?
How does $\tilde{A}_\rho$ behave over time scales
during which $\C(t)$ exhibits known behaviors
(e.g., until the dissipation time
or from the dissipation time to the scrambling time~\cite{Swingle_16_Measuring})?
Under what conditions does $\tilde{A}_\rho$ behave nonclassically
(assume negative or nonreal values)?
How does a chaotic system's $\tilde{A}_\rho$ look?
These questions are under investigation~\cite{BrianDisc}.

As another example, fluctuation relations have been used
to estimate the free-energy difference $\Delta F$
from experimental data.
Experimental measurements of $\C(t)$ are possible for certain platforms,
in certain regimes~\cite{Swingle_16_Measuring,Yao_16_Interferometric,Zhu_16_Measurement,Li_16_Measuring,Garttner_16_Measuring}.
Theorem~\ref{theorem:OTOC_FT} expands the set of platforms and regimes.
Measuring quantum amplitudes, as via weak measurements~\cite{Lundeen_11_Direct,Lundeen_12_Procedure,Bamber_14_Observing,Mirhosseini_14_Compressive},
now offers access to $\C(t)$.
Inferring small systems' $\tilde{A}_\rho$'s with existing platforms~\cite{White_16_Preserving}
might offer a challenge for the near future.

Finally, Theorem~\ref{theorem:OTOC_FT} can provide 
a new route to bounding $\C(t)$.
A Lyapunov exponent $\lambda_{\rm L}$ governs the chaotic decay of $\C(t)$.
The exponent has been bounded, 
including with Lieb-Robinson bounds and complex analysis~\cite{Maldacena_15_Bound,Lashkari_13_Towards,Kitaev_15_Simple}.
The right-hand side of Eq.~\eqref{eq:Result} can provide
an independent bounding method that offers new insights.

%
%
\section*{Acknowledgements}
I have the pleasure of thanking Fernando G. S. L. Brand\~{a}o, Jordan S. Cotler, David Ding, Yichen Huang, Geoff Penington, Evgeny Mozgunov, Brian Swingle, Christopher D. White, and Yong-Liang Zhang for valuable discussions. I am grateful to Justin Dressel for sharing weak-measurement and quasiprobability expertise, to Sebastian Deffner and John Goold for discussing fluctuation relations' forms, to Alexey Gorshkov for sharing interference expertise, to Alexei Kitaev for introducing me to the OTOC during Ph 219c preparations, to John Preskill for suggesting combining the OTOC with fluctuation relations, and to Norman Yao for spotlighting degeneracies.
This research was supported by NSF grant PHY-0803371. The Institute for Quantum Information and Matter (IQIM) is an NSF Physics Frontiers Center supported by the Gordon and Betty Moore Foundation.

\begin{appendices}

\renewcommand{\thesubsection}{\Alph{section}.\arabic{subsection}}
\renewcommand{\thesubsection}{\Alph{section}.\arabic{subsection}}
\renewcommand{\thesubsubsection}{\Alph{section}.\arabic{subsection}.\roman{subsubsection}}

\makeatletter\@addtoreset{equation}{section}
\def\theequation{\thesection\arabic{equation}}

%
%
\section{Weak measurement of 
the combined quantum amplitude $\tilde{A}_\rho$}
\label{section:WeakMeas}

$\tilde{A}_\rho$ [Eq.~\eqref{eq:TildeAExp}] 
resembles the Kirkwood-Dirac quasiprobability
for a quantum state~\cite{Kirkwood_33_Quantum,Dirac_45_On,Dressel_15_Weak}.
This quasiprobability has been inferred
from weak-measurement experiments~\cite{Lundeen_11_Direct,Lundeen_12_Procedure,Bamber_14_Observing,Mirhosseini_14_Compressive,Dressel_14_Understanding}.
Weak measurements have been performed on cold atoms~\cite{Smith_04_Continuous},
which have been proposed as platforms 
for realizing scrambling and quantum chaos~\cite{Swingle_16_Measuring,Yao_16_Interferometric,Danshita_16_Creating}.

$\tilde{A}_\rho$ can be inferred from many instances of a protocol $\Protocol_\weak$.
$\Protocol_\weak$ consists of a state preparation,
three evolutions interleaved with three weak measurements,
and a strong measurement.
The steps appear in Sec.~\ref{section:WeakMain}. 

I here flesh out the protocol,
assuming that the system, $S$, begins in
the infinite-temperature Gibbs state: $\rho = \id / \Dim$.
$\tilde{A}_\rho$ simplifies as in Eq.~\eqref{eq:TildeASimple}.
The final factor becomes $p_{w_3, \DegenW_{w_3} }  =  1 / \Dim$.
The number of weak measurements in $\Protocol_\weak$ reduces to two.
Generalizing to arbitrary $\rho$'s
is straightforward but requires lengthier calculations
and more ``background'' terms.

Each trial in the simplified $\Protocol_\weak$ consists of a state preparation,
three evolutions interleaved with two weak measurements,
and a strong measurement.
Loosely, one performs the following protocol:
Prepare $\ket{ w_3,  \DegenW_{w_3} }$.
Evolve $S$ backward under $U^\dag$.
Measure $\ketbra{ v_1,  \DegenV_{v_1} }{ v_1,  \DegenV_{v_1} }$ weakly.
Evolve $S$ forward under $U$.
Measure $\ketbra{ w_2,  \DegenW_{w_2} }{ w_2,  \DegenW_{w_2} }$ weakly.
Evolve $S$ backward under $U^\dag$.
Measure $\ketbra{ v_2,  \DegenV_{v_2} }{ v_2,  \DegenV_{v_2} }$ strongly.

Let us analyze the protocol in greater detail.
The $\ket{ w_3,  \DegenW_{w_3} }$ preparation and backward evolution yield
$\ket{ \psi }  =  U^\dag \ket{ w_3,  \DegenW_{w_3} }$.
The weak measurement of 
$\ketbra{ v_1,  \DegenV_{v_1} }{ v_1,  \DegenV_{v_1} }$ is implemented as follows:
$S$ is coupled weakly to an ancilla $\mathcal{A}_a$.
The observable $\NondegV$ of $S$ comes to be correlated
with an observable of $\mathcal{A}_a$.
Example $\mathcal{A}_a$ observables include
a pointer's position on a dial
and a component $\sigma_\ell$ of a qubit's spin (wherein $\ell = x, y, z$).
The $\mathcal{A}_a$ observable is measured projectively.
Let $x$ denote the measurement's outcome.
$x$ encodes partial information about the system's state.
We label by $( v_1,  \DegenV_{v_1} )$ the $\NondegV$ eigenvalue
most reasonably attributable to $S$
if the $\mathcal{A}_a$ measurement yields $x$.

The coupling and the $\mathcal{A}_a$ measurement
evolve $\ket{ \psi }$ under the Kraus operator~\cite{NielsenC10}
\begin{align}
   \label{eq:Mx}
   M_x  =  \sqrt{p_a(x)} \: \id
   + g_a(x)  \,  \ketbra{ v_1,  \DegenV_{v_1} }{ v_1,  \DegenV_{v_1} }   \, .
\end{align}
Equation~\eqref{eq:Mx} can be derived, e.g., from
the Gaussian-meter model~\cite{Dressel_15_Weak,Aharonov_88_How}
or the qubit-meter model~\cite{White_16_Preserving}.
The projector can be generalized to 
a projector $\Pi_{v_1}$ onto a degenerate eigensubspace.
The generalization may decrease exponentially
the number of trials required~\cite{BrianDisc}.
By the probabilistic interpretation of quantum channels,
the baseline probability $p_a(x)$ denotes
the likelihood that, in any given trial,
$S$ fails to couple to $\mathcal{A}_a$
but the $\mathcal{A}_a$ measurement yields $x$ nonetheless.
The detector is assumed, for convenience, to be calibrated such that 
\begin{align}
   \label{eq:Calibrate}
   \int dx \cdot x  \: p_a(x)  =  0 \, .
\end{align}
The small tunable parameter $g_a(x)$ 
quantifies the coupling strength.

The system's state becomes 
$\ket{ \psi' }  =  M_x U^\dag \ket{ w_3,  \DegenW_{w_3} }$,
to within a normalization factor.
$S$ evolves under $U$ as
\begin{align}
  \ket{ \psi' }  \mapsto  \ket{ \psi'' }
   =  U M_x U^\dag \ket{ w_3,  \DegenW_{w_3} } \, ,
\end{align}
to within normalization.
$\ketbra{ w_2,  \DegenW_{w_2} }{ w_2,  \DegenW_{w_2} }$ is measured weakly:
$S$ is coupled weakly to an ancilla $\mathcal{A}_b$.
$\NondegW$ comes to be correlated with
a pointer-like variable of $\mathcal{A}_b$.
The pointer-like variable is measured projectively.
Let $y$ denote the outcome.
The coupling and measurement evolve $\ket{ \psi'' }$ 
under the Kraus operator
\begin{align}
   \label{eq:My}
   M_y  =  \sqrt{ p_b(y) }  \:  \id
   +  g_b(y)  \,  \ketbra{ w_2,  \DegenW_{w_2} }{ w_2,  \DegenW_{w_2} }  \, .
\end{align}
The system's state becomes
$\ket{ \psi''' }  =  M_y U M_x U^\dag \ket{ w_3,  \DegenW_{w_3} } $,
to within normalization.
The state evolves backward under $U^\dag$.
Finally, $\NondegV$ is measured projectively.

Each trial involves two weak measurements
and one strong measurement.
The probability that the measurements yield the outcomes $x$, $y$, 
and $( v_2,  \DegenV_{v_2} )$ is
\begin{align}
   \label{eq:PWeak}
   \PWeak \LParen x, y,  ( v_2,  \DegenV_{v_2} ) \RParen =  
   | \langle v_2,  \DegenV_{v_2} | U^\dag M_y U M_x U^\dag 
   | w_3,  \DegenW_{w_3} \rangle |^2 \, .
\end{align}
Integrating over $x$ and $y$ yields
\begin{align}
   \label{eq:WeakInt}
   \WeakInt  :=
   \int dx \; dy \cdot x \, y \; \PWeak \LParen x,y, ( v_2,  \DegenV_{v_2} ) \RParen \, .
\end{align}
We substitute in for $M_x$ and $M_y$ 
from Eqs.~\eqref{eq:Mx} and~\eqref{eq:My},
then multiply out.
We approximate to second order in the weak-coupling parameters.
The calibration condition~\eqref{eq:Calibrate}
causes terms to vanish:
\begin{align}
   \label{eq:WeakInt2}
   \WeakInt  &  =  \int dx \; dy  \cdot  x \:  y  \;   \sqrt{ p_a(x) \: p_b(y)  }  
   \Big[  g_a (x) \, g_b(y)  \cdot \Dim
            \nonumber \\ &    \times  
            \tilde{A}_{\id / \Dim} ( w, v, \DegenW_w, \DegenV_v )
            +  \cc \Big] 
   +  \int dx \; dy  \cdot  x \:  y  \;
   \sqrt{ p_a(x) \: p_b(y)  }
   \nonumber \\ &  \times
   \Big[  g_a^*(x) \, g_b(y)  \,  
   \langle v_2,  \DegenV_{v_2} | U^\dag |  w_2,  \DegenW_{w_2}  \rangle 
   \langle  w_2,  \DegenW_{w_2}  | w_3,  \DegenW_{w_3} \rangle
   \nonumber \\ &  \times
   \LParen \langle v_2,  \DegenV_{v_2} | v_1,  \DegenV_{v_1} \rangle 
   \langle v_1,  \DegenV_{v_1} | U^\dag 
   | w_3,  \DegenW_{w_3} \rangle \RParen^*
   + \cc \Big] 
   \nonumber \\ & 
   + O \LParen g_a(x)^2 \, g_b(y) \RParen
   + O \LParen g_a(x) \, g_b(y)^2 \RParen \, .
\end{align}

The baseline probabilities $p_a(x)$ and $p_b(x)$
are measured during calibration.
Let us focus on the second integral.
By orthonormality, $\langle  w_2,  \DegenW_{w_2}  | w_3,  \DegenW_{w_3} \rangle 
= \delta_{ w_2 w_3}  \,  \delta_{ \DegenW_{w_2}  \DegenW_{w_3} }$,
and $\langle v_2,  \DegenV_{v_2} |  v_1,  \DegenV_{v_1}  \rangle  
=  \delta_{v_2  v_1}  \,  \delta_{ \DegenV_{v_2}  \DegenV_{v_1} }$.
The integral vanishes if 
$( w_3,  \DegenW_{w_3} ) \neq ( w_2,  \DegenW_{w_2} )$ 
or if $( v_2,  \DegenV_{v_2} ) \neq  ( v_1,  \DegenV_{v_1} )$.
Suppose that $( w_3,  \DegenW_{w_3} ) = ( w_2,  \DegenW_{w_2} )$ 
and $( v_2,  \DegenV_{v_2} ) = ( v_1,  \DegenV_{v_1} )$.
The second integral becomes
\begin{align}
   & \int dx \; dy  \cdot  x \:  y  \;
   \sqrt{ p_a(x) \, p_b(y) }  \:  
   \Big[ g_a^*(x) \, g_b(y) \,
   \nonumber \\ & \qquad \times
   | \langle v_2,  \DegenV_{v_2} | U^\dag |  w_3,  \DegenW_{w_3}  \rangle |^2
   + \cc \Big]  .
\end{align}
The square modulus, a probability,
can be measured via Born's rule.
The experimenter controls $g_a(x)$ and $g_b(y)$.
The second integral in Eq.~\eqref{eq:WeakInt2}
is therefore known.

From the first integral, we infer about   $\tilde{A}_{\id / \Dim}$.
Consider trials in which the couplings are chosen such that
\begin{align}
   \alpha  :=  
    \int dx \; dy  \cdot  x \:  y  \;
   \sqrt{ p_a(x) \, p_b(y) }  \:  
   g_a(x) \, g_b(y) \in \mathbb{R} \, .
\end{align}
The first integral becomes
$2 \alpha \, \Dim  \, \Re \LParen 
\tilde{A}_{\id / \Dim}  ( w, v, \DegenW_w, \DegenV_v ) \RParen$.
From these trials, one infers the real part of $\tilde{A}_{\id / \Dim}$.
Now, consider trials in which
$i \, \alpha  \in  \mathbb{R}$.
The first bracketed term becomes
$2 | \alpha | \, \Dim  \,  \Im \LParen 
\tilde{A}_{\id / \Dim}  ( w, v, \DegenW_w, \DegenV_v ) \RParen \, .$
From these trials, one infers the imaginary part of $\tilde{A}_{\id / \Dim}$.

$\alpha$ can be tuned between real and imaginary in practice~\cite{Lundeen_11_Direct}.
Consider a weak measurement in which the ancillas are qubits.
An ancilla's $\sigma_y$ can be coupled to a system observable.
Whether the ancilla's $\sigma_x$ or $\sigma_y$ is measured
dictates whether $\alpha$ is real or imaginary.

The combined quantum amplitude $\tilde{A}_\rho$ 
can therefore be inferred from weak measurements.
$\tilde{A}_\rho$ can be measured alternatively via interference.

%
%
\section{Interference-based measurement of the combined quantum amplitude
$\tilde{A}_\rho$}
\label{section:Interfere} 

I detail an interference-based scheme for measuring 
$\tilde{A}_\rho( w, v, \DegenW_w, \DegenV_v )$ 
[Eq.~\eqref{eq:TildeAExp}].
The scheme requires no reversal of the time evolution in any trial.
As implementing time reversal can be difficult,
the absence of time reversal can benefit 
OTOC-measurement schemes~\cite{Yao_16_Interferometric,Zhu_16_Measurement}.

I specify how to measure an inner product
$z  :=  \langle a | \U | b \rangle$,
wherein $a, b  \in \{  ( w_\ell,  \DegenW_{w_\ell } ),
 ( v_m ,  \DegenV_{v_m}  )  \}$ 
 and $\U  \in  \{ U, U^\dag \}$.
Then, I discuss measurements of the state-dependent factor
in Eq.~\eqref{eq:TildeAExp}.

The inner product $z$ is measured as follows.
The system $S$ is initialized 
to some fiducial state $\ket{f}$.
An ancilla qubit $\mathcal{A}$ is prepared in the state 
$\frac{1}{\sqrt{2} }( \ket{0} + \ket{1} )$.
The $+1$ and $-1$ eigenstates of $\sigma_z$
are denoted by $\ket{0}$ and $\ket{1}$.
The composite system $\mathcal{A}S$ begins in the state
$\ket{ \psi }  =  \frac{1}{\sqrt{2} } ( \ket{0} \ket{f}  +  \ket{1} \ket{f} )$.

A unitary is performed on $S$, conditioned on $\mathcal{A}$:
If $\mathcal{A}$ is in state $\ket{0}$, 
then $S$ is brought to state $\ket{b}$,
and $\U$ is applied to $S$.
If $\mathcal{A}$ is in state $\ket{1}$,
$S$ is brought to state $\ket{a}$.
The global state becomes
$\ket{ \psi' }  =  \frac{1}{ \sqrt{2} } [
   \ket{0} ( \U \ket{b} )  +  \ket{1} \ket{a} ) ] \, .$
A unitary $e^{ - i \theta \sigma_x }$
rotates the ancilla's state 
through an angle $\theta$ about the $x$-axis.
The global state becomes
\begin{align}
   \label{eq:PsiPP}
   \ket{ \psi'' }  & =  \frac{1}{ \sqrt{2} }  \Bigg[
   \left(  \cos  \frac{\theta}{2}  \ket{0}  
   -  i \sin \frac{ \theta }{ 2 }  \ket{1}   \right)  
   ( \U \ket{ b }  )
   \nonumber \\ & \qquad
   +  \left(  - i \sin \frac{ \theta }{ 2 }  \ket{0}  
   +  \cos  \frac{ \theta }{ 2 }  \ket{1} \right)
   \ket{a }  \Bigg]  \, .
\end{align}

The ancilla's $\sigma_z$ is measured,
and the system's $\{ \ket{ a } \}$ is measured.
The probability that the measurements
yield $+1$ and $a$ is
\begin{align} 
   \Prob(+1, a)  & =  \frac{1}{4} ( 1 - \sin \theta)
   \Bigg(  \cos^2  \frac{ \theta }{ 2 }  \:  | z |^2
   -  \sin \theta  \:  \Im  (z)
   +  \sin^2 \frac{ \theta }{ 2 }  \Bigg) \, .
\end{align}
The imaginary part of $z$ is denoted by $\Im(z)$.
$\Prob(+1, a)$ can be inferred from the outcomes of multiple trials.
The $| z |^2$, representing a probability, can be measured independently.
From the $| z |^2$ and $\Prob(+1, a)$ measurements,
$\Im (z)$ can be inferred.

$\Re(z)$ can be inferred from another set of interference experiments.
The rotation about $\hat{x}$ is replaced with
a rotation about $\hat{y}$.
The unitary $e^{-i \phi \sigma_y }$ implements this rotation,
through an angle $\phi$.
Equation~\eqref{eq:PsiPP} becomes
\begin{align}
   \ket{ \tilde{ \psi}'' } & =
   \frac{1}{ \sqrt{2} }  \Big[  
   \left(  \cos \frac{ \phi }{ 2 }  \,  \ket{0}  +  \sin \frac{ \phi }{2} \, \ket{1}  \right)
   ( \U \, \ket{b} )
   \nonumber \\ & \qquad \qquad +  
   \left(  - \sin \frac{\phi }{2}  \,  \ket{0}
   +  \cos \frac{ \phi }{2}  \,  \ket{1}  \right)  \ket{a}  
   \Big] \, .
\end{align}
The ancilla's $\sigma_z$ and the system's $\{ \ket{a} \}$
are measured.
The probability that the measurements yield $+1$ and $a$ is
\begin{align}
   \TProb(+1, a)  & =  \frac{1}{4}  ( 1 - \sin \phi )
   \Bigg(  \cos^2 \frac{\phi}{2}  \,  | z |^2
   \nonumber \\ & \qquad
   - \sin \phi \; \Re(z)  +  \sin^2  \frac{\phi}{2}  \Bigg) \, .
\end{align}
One measures $\TProb(+1, a)$ and $|z|^2$,
then infers  $\Re(z)$.
The real and imaginary parts of $z$ are thereby gleaned
from interferometry.

Equation~\eqref{eq:TildeAExp} contains the state-dependent factor
$M := \langle v_1,  \DegenV_{v_1}  |  
  \rho  U^\dag  |  w_3,  \DegenW_{w_3}  \rangle$.
This factor is measured easily if 
$\rho$ shares its eigenbasis with $\NondegW(t)$ or with $\NondegV$.
In these cases, $M$ assumes the form 
$\langle a | U^\dag | b \rangle \, p$.
The inner product is measured as above.
The probability $p$ is measured via Born's rule.
In an important subcase, $\rho$ is 
the infinite-temperature Gibbs state $\id / \Dim$.
The system's size sets $p = 1 / \Dim$.
Outside of these cases, $M$ can be inferred from quantum tomography~\cite{Paris_04_Q_State_Estimation}.
Tomography requires many trials but is possible in principle
and can be realized with small systems.

%

\end{appendices}

%
%
\bibliographystyle{h-physrev}
\bibliography{OTOC_FT_bib}

\begin{thebibliography}{10}

\bibitem{Shenker_Stanford_14_BHs_and_butterfly}
S.~H. {Shenker} and D.~{Stanford},
\newblock Journal of High Energy Physics {\bf 3}, 67 (2014).

\bibitem{Shenker_Stanford_14_Multiple_shocks}
S.~H. {Shenker} and D.~{Stanford},
\newblock Journal of High Energy Physics {\bf 12}, 46 (2014).

\bibitem{Shenker_Stanford_15_Stringy}
S.~H. {Shenker} and D.~{Stanford},
\newblock Journal of High Energy Physics {\bf 5}, 132 (2015).

\bibitem{Roberts_15_Localized_shocks}
D.~A. {Roberts}, D.~{Stanford}, and L.~{Susskind},
\newblock Journal of High Energy Physics {\bf 3}, 51 (2015).

\bibitem{Roberts_Stanford_15_Diagnosing}
D.~A. {Roberts} and D.~{Stanford},
\newblock Physical Review Letters {\bf 115}, 131603 (2015).

\bibitem{Maldacena_15_Bound}
J.~{Maldacena}, S.~H. {Shenker}, and D.~{Stanford},
\newblock ArXiv e-prints  (2015), 1503.01409.

\bibitem{LarkinO_69}
A.~Larkin and Y.~N. Ovchinnikov,
\newblock Soviet Journal of Experimental and Theoretical Physics {\bf 28}
  (1969).

\bibitem{Sekino_Susskind_08_Fast_scramblers}
Y.~Sekino and L.~Susskind,
\newblock Journal of High Energy Physics {\bf 2008}, 065 (2008).

\bibitem{Huang_16_MBL_OTOC}
Y.~{Huang}, Y.-L. {Zhang}, and X.~{Chen},
\newblock ArXiv e-prints  (2016), 1608.01091.

\bibitem{Swingle_16_MBL_OTOC}
B.~{Swingle} and D.~{Chowdhury},
\newblock ArXiv e-prints  (2016), 1608.03280.

\bibitem{Fan_16_MBL_OTOC}
R.~{Fan}, P.~{Zhang}, H.~{Shen}, and H.~{Zhai},
\newblock ArXiv e-prints  (2016), 1608.01914.

\bibitem{He_16_MBL_OTOC}
R.-Q. {He} and Z.-Y. {Lu},
\newblock ArXiv e-prints  (2016), 1608.03586.

\bibitem{Chen_16_MBL_OTOC}
Y.~{Chen},
\newblock ArXiv e-prints  (2016), 1608.02765.

\bibitem{HosurYoshida_16_Chaos}
P.~{Hosur}, X.-L. {Qi}, D.~A. {Roberts}, and B.~{Yoshida},
\newblock Journal of High Energy Physics {\bf 2}, 4 (2016), 1511.04021.

\bibitem{Swingle_16_Measuring}
B.~{Swingle}, G.~{Bentsen}, M.~{Schleier-Smith}, and P.~{Hayden},
\newblock ArXiv e-prints  (2016), 1602.06271.

\bibitem{Yao_16_Interferometric}
N.~Y. {Yao} {\em et~al.},
\newblock ArXiv e-prints  (2016), 1607.01801.

\bibitem{Zhu_16_Measurement}
G.~{Zhu}, M.~{Hafezi}, and T.~{Grover},
\newblock ArXiv e-prints  (2016), 1607.00079.

\bibitem{Li_16_Measuring}
J.~{Li} {\em et~al.},
\newblock ArXiv e-prints  (2016), 1609.01246.

\bibitem{Garttner_16_Measuring}
M.~{G{\"a}rttner} {\em et~al.},
\newblock ArXiv e-prints  (2016), 1608.08938.

\bibitem{Maldacena_16_Comments}
J.~{Maldacena} and D.~{Stanford},
\newblock ArXiv e-prints  (2016), 1604.07818.

\bibitem{Polchinski_16_Spectrum}
J.~{Polchinski} and V.~{Rosenhaus},
\newblock Journal of High Energy Physics {\bf 4}, 1 (2016), 1601.06768.

\bibitem{Jarzynski97}
C.~Jarzynski,
\newblock Physical Review Letters {\bf 78}, 2690 (1997).

\bibitem{Crooks99}
G.~E. Crooks,
\newblock Physical Review E {\bf 60}, 2721 (1999).

\bibitem{Tasaki00}
H.~Tasaki,
\newblock arXiv e-print  (2000), cond-mat/0009244.

\bibitem{Kurchan00}
J.~{Kurchan},
\newblock eprint arXiv:cond-mat/0007360  (2000), cond-mat/0007360.

\bibitem{Saira_12_Test}
O.-P. Saira {\em et~al.},
\newblock Phys. Rev. Lett. {\bf 109}, 180601 (2012).

\bibitem{Chipot_07_Free}
C.~Chipot and A.~Pohorille, editors,
\newblock {\em Free Energy Calculations: Theory and Applications in Chemistry
  and Biology}, Springer Series in Chemical Physics Vol.~86 (Springer-Verlag,
  2007).

\bibitem{CollinRJSTB05}
D.~Collin {\em et~al.},
\newblock Nature {\bf 437}, 231 (2005).

\bibitem{Douarche_05_Experimental}
F.~Douarche, S.~Ciliberto, A.~Petrosyan, and I.~Rabbiosi,
\newblock EPL (Europhysics Letters) {\bf 70}, 593 (2005).

\bibitem{Blickle_06_Thermo}
V.~Blickle, T.~Speck, L.~Helden, U.~Seifert, and C.~Bechinger,
\newblock Phys. Rev. Lett. {\bf 96}, 070603 (2006).

\bibitem{Harris_07_Experimental}
N.~C. Harris, Y.~Song, and C.-H. Kiang,
\newblock Phys. Rev. Lett. {\bf 99}, 068101 (2007).

\bibitem{MossaMFHR09}
A.~Mossa, M.~Manosas, N.~Forns, J.~M. Huguet, and F.~Ritort,
\newblock Journal of Statistical Mechanics: Theory and Experiment {\bf 2009},
  P02060 (2009).

\bibitem{ManosasMFHR09}
M.~{Manosas}, A.~{Mossa}, N.~{Forns}, J.~M. {Huguet}, and F.~{Ritort},
\newblock Journal of Statistical Mechanics: Theory and Experiment {\bf 2009},
  P02061 (2009).

\bibitem{Batalhao_14_Experimental}
T.~B. {Batalh{\~a}o} {\em et~al.},
\newblock Physical Review Letters {\bf 113}, 140601 (2014), 1308.3241.

\bibitem{An_15_Experimental}
S.~{An} {\em et~al.},
\newblock Nature Physics {\bf 11}, 193 (2015).

\bibitem{Maruyama_09_Colloquium}
K.~Maruyama, F.~Nori, and V.~Vedral,
\newblock Rev. Mod. Phys. {\bf 81}, 1 (2009).

\bibitem{Aberg_13_Truly}
J.~{{\AA}berg},
\newblock Nature Communications {\bf 4}, 1925 (2013), 1110.6121.

\bibitem{YungerHalpern_15_Introducing}
N.~Yunger~Halpern, A.~J.~P. Garner, O.~C.~O. Dahlsten, and V.~Vedral,
\newblock New Journal of Physics {\bf 17}, 095003 (2015).

\bibitem{Salek_15_Fluctuations}
S.~{Salek} and K.~{Wiesner},
\newblock ArXiv e-prints  (2015), 1504.05111.

\bibitem{YungerHalpern_15_What}
N.~{Yunger Halpern}, A.~J.~P. {Garner}, O.~C.~O. {Dahlsten}, and V.~{Vedral},
\newblock ArXiv e-prints  (2015), 1505.06217.

\bibitem{Dahlsten_15_Equality}
O.~{Dahlsten} {\em et~al.},
\newblock ArXiv e-prints  (2015), 1504.05152.

\bibitem{Alhambra_16_Fluctuating_Work}
A.~M. Alhambra, L.~Masanes, J.~Oppenheim, and C.~Perry,
\newblock Phys. Rev. X {\bf 6}, 041017 (2016).

\bibitem{BrianDisc}
J.~Dressel, B.~Swingle, and N.~Yunger~Halpern,
\newblock (in prep).

\bibitem{TLH_07_Work}
P.~Talkner, E.~Lutz, and P.~H\"anggi,
\newblock Phys. Rev. E {\bf 75}, 050102 (2007).

\bibitem{Kirkwood_33_Quantum}
J.~G. Kirkwood,
\newblock Physical Review {\bf 44}, 31 (1933).

\bibitem{Dirac_45_On}
P.~A.~M. Dirac,
\newblock Reviews of Modern Physics {\bf 17}, 195 (1945).

\bibitem{Dressel_15_Weak}
J.~Dressel,
\newblock Phys. Rev. A {\bf 91}, 032116 (2015).

\bibitem{Carmichael_02_Statistical}
H.~J. Carmichael,
\newblock {\em Statistical Methods in Quantum Optics I: Master Equations and
  Fokker-Planck Equations} (Springer-Verlag, 2002).

\bibitem{Ferrie_11_Quasi}
C.~{Ferrie},
\newblock Reports on Progress in Physics {\bf 74}, 116001 (2011).

\bibitem{Lundeen_11_Direct}
J.~S. Lundeen, B.~Sutherland, A.~Patel, C.~Stewart, and C.~Bamber,
\newblock Nature {\bf 474}, 188 (2011).

\bibitem{Lundeen_12_Procedure}
J.~S. Lundeen and C.~Bamber,
\newblock Phys. Rev. Lett. {\bf 108}, 070402 (2012).

\bibitem{Bamber_14_Observing}
C.~Bamber and J.~S. Lundeen,
\newblock Phys. Rev. Lett. {\bf 112}, 070405 (2014).

\bibitem{Mirhosseini_14_Compressive}
M.~Mirhosseini, O.~S. Maga\~na Loaiza, S.~M. Hashemi~Rafsanjani, and R.~W.
  Boyd,
\newblock Phys. Rev. Lett. {\bf 113}, 090402 (2014).

\bibitem{Dressel_14_Understanding}
J.~Dressel, M.~Malik, F.~M. Miatto, A.~N. Jordan, and R.~W. Boyd,
\newblock Rev. Mod. Phys. {\bf 86}, 307 (2014).

\bibitem{Kofman_12_Nonperturbative}
A.~G. Kofman, S.~Ashhab, and F.~Nori,
\newblock Physics Reports {\bf 520}, 43  (2012),
\newblock Nonperturbative theory of weak pre- and post-selected measurements.

\bibitem{White_16_Preserving}
T.~C. White {\em et~al.},
\newblock npj Quantum Information {\bf 2} (2016).

\bibitem{Dressel_14_Implementing}
J.~Dressel, T.~A. Brun, and A.~N. Korotkov,
\newblock Phys. Rev. A {\bf 90}, 032302 (2014).

\bibitem{Smith_04_Continuous}
G.~A. Smith, S.~Chaudhury, A.~Silberfarb, I.~H. Deutsch, and P.~S. Jessen,
\newblock Phys. Rev. Lett. {\bf 93}, 163602 (2004).

\bibitem{Danshita_16_Creating}
I.~{Danshita}, M.~{Hanada}, and M.~{Tezuka},
\newblock ArXiv e-prints  (2016), 1606.02454.

\bibitem{Lashkari_13_Towards}
N.~Lashkari, D.~Stanford, M.~Hastings, T.~Osborne, and P.~Hayden,
\newblock Journal of High Energy Physics {\bf 2013}, 22 (2013).

\bibitem{Kitaev_15_Simple}
A.~Kitaev,
\newblock A simple model of quantum holography, 2015.

\bibitem{NielsenC10}
M.~A. Nielsen and I.~L. Chuang,
\newblock {\em {Quantum Computation and Quantum Information}} (Cambridge
  University Press, 2010).

\bibitem{Aharonov_88_How}
Y.~Aharonov, D.~Z. Albert, and L.~Vaidman,
\newblock Phys. Rev. Lett. {\bf 60}, 1351 (1988).

\bibitem{Paris_04_Q_State_Estimation}
M.~Paris and J.~Rehacek, editors,
\newblock {\em {Quantum State Estimation}}, Lecture Notes in Physics Vol. 649
  (Springer, Berlin, Heidelberg, 2004).

\end{thebibliography}


\end{document}